\documentclass[prd,aps,twocolumn,preprintnumbers, showpacs,
nofootinbib,superscriptaddress,notitlepage,floatfix]{revtex4-1}
\usepackage{amsthm,amsmath,amssymb}
\usepackage{mathrsfs}
\usepackage{graphicx}
\usepackage{slashed}
\usepackage[colorlinks, citecolor=blue,anchorcolor=blue,menucolor=blue,
linkcolor=blue,filecolor=blue,runcolor=blue,urlcolor=blue,frenchlinks=true]{hyperref}
\usepackage[normalem]{ulem}

\begin{document}

\title{Gluon Parton Distribution of the Pion from Lattice QCD }
\author{Zhouyou Fan}
\affiliation{Department of Physics and Astronomy, Michigan State University, East Lansing, MI 48824}

\author{Huey-Wen Lin}
\affiliation{Department of Physics and Astronomy, Michigan State University, East Lansing, MI 48824}
\affiliation{Department of Computational Mathematics,
  Science and Engineering, Michigan State University, East Lansing, MI 48824}

\preprint{MSUHEP-21-004}

\begin{abstract}
We present the first determination of the $x$-dependent pion gluon distribution from lattice QCD using the pseudo-PDF approach.
We use lattice ensembles with 2+1+1 flavors of highly improved staggered quarks (HISQ), generated by MILC Collaboration, at two lattice spacings $a\approx 0.12$ and 0.15~fm and three pion masses $M_\pi\approx 220$, 310 and 690~MeV.
We use clover fermions for the valence action and momentum smearing to achieve pion boost momentum up to 2.29~GeV.
We find that the dependence of the pion gluon parton distribution on lattice spacing and pion mass is mild.
We compare our results from the lightest pion mass ensemble with the determination by JAM and xFitter global fits.
\end{abstract}

\maketitle

\section{Introduction}

The lightest bound state in quantum chromodynamics (QCD), the pion, plays a fundamental role, since it is the Nambu-Goldstone boson of dynamical chiral symmetry breaking (DCSB).
Studies of pion and kaon structure reveal the physics of DCSB, help to reveal the relative impact of DCSB versus the chiral symmetry breaking by the quark masses, and are important to understand nonperturbative QCD.
Studying the pion parton distribution functions (PDFs) is important to characterize the structure of the pion and further understand DCSB and nonperturbative QCD.
Currently, our knowledge of the pion PDFs is less than the nucleon PDFs, because there are fewer experimental data sets, especially for the sea-quark and gluon distributions.
The future U.S.-based Electron-Ion Collider (EIC)~\cite{Accardi:2012qut}, planned to be built at Brookhaven National Lab, will further our knowledge of  pion structure~\cite{Arrington:2021biu,Aguilar:2019teb}.
In China, a similar machine, the Electron-Ion Collider in China (EicC)~\cite{Anderle:2021wcy}, is also planned to make impacts on the pion gluon and sea-quark distributions.
In Europe, the Drell-Yan and $J/\psi$-production experiments from COMPASS++/AMBER~\cite{Denisov:2018unj} will aim at improving our knowledge of both the pion gluon and quark PDFs.

Global analyses of pion PDFs mostly rely on Drell-Yan data.
The early studies of pion PDFs were based mostly on pion-induced Drell-Yan data and use $J/\psi$-production data or direct photon production to constrain the pion gluon PDF~\cite{Owens:1984zj,Aurenche:1989sx,Sutton:1991ay,Gluck:1991ey,Gluck:1999xe}.
There are more recent studies, such as the work by Bourrely and Soffer~\cite{Bourrely:2018yck}, that extract the pion PDF based on Drell-Yan $\pi^+W$ data.
JAM Collaboration~\cite{Barry:2018ort,Cao:2021aci} uses a Monte-Carlo approach to analyze the Drell-Yan $\pi A$ and leading-neutron electroproduction data from HERA to reach the lower-$x$ region, and revealed that gluons carry a significantly higher momentum fraction (about $30\%$) in the pion than had been inferred from Drell-Yan data alone.
The xFitter group~\cite{Novikov:2020snp} analyzed Drell-Yan $\pi A$ and photoproduction data using their open-source QCD fit framework for PDF extraction and found that these data can constrain the valence distribution well but are not sensitive enough for the sea and gluon distributions to be precisely determined.
The analysis done Ref.~\cite{Chang:2020rdy} suggests that the pion-induced $J/\psi$-production data has additional constraint on pion PDFs, particularly in the pion gluon PDF in the large-$x$ region.
All in all, the pion valence-quark distributions are better constrained
than the gluon distribution from the global analysis of experimental data.
While waiting for more experimental data sets, the study of the pion gluon distribution from theoretical side can provide useful information for the experiments.

The pion gluon PDF is rarely studied using continuum-QCD phenomenological models or through lattice-QCD (LQCD) simulations.
Most model studies only predict the pion valence-quark distribution~\cite{Nam:2012vm,Watanabe:2016lto,Watanabe:2017pvl,Hutauruk:2016sug,Lan:2019vui,Lan:2019rba,deTeramond:2018ecg,Watanabe:2019zny,Han:2018wsw,Chang:2014lva,Chang:2014gga,Chen:2016sno,Shi:2018mcb,Bednar:2018mtf,Ding:2019lwe},
but the gluon and sea PDFs are predicted by the Dyson-Schwinger equation (DSE) continuum approach~\cite{Freese:2021zne}. The prediction of the pion gluon PDF in DSE, based on an implementation of rainbow-ladder truncation of DSE, is consistent with the JAM pion gluon PDF result~\cite{Barry:2018ort,Cao:2021aci} within two sigma.
LQCD provides an first-principles calculations to improve our knowledge of nonperturbative pion gluon structure;
however, there have been only two efforts to determine the first moment of pion gluon PDF~\cite{Meyer:2007tm,Shanahan:2018pib}.
An early calculation in 2000 using quenched QCD predicted $\langle x \rangle_g=0.37(8)(12)$, using Wilson fermion action with a lattice spacing $a=0.093$~fm, lattice size $L^3\times T=24^4$, a large 890-MeV pion mass and 3,066 configurations at $\mu^2=4\text{ GeV}^2$~\cite{Meyer:2007tm}.
A more recent study in 2018 using $N_f=2+1$ clover fermion action  with a lattice spacing $a=0.1167(16)$~fm, larger lattice size $32^3\times 96$, 450-MeV pion mass, and 572,663 measurements, gave a larger first-moment result, $\langle x \rangle_g=0.61(9)$ at $\mu^2=4\text{ GeV}^2$~\cite{Shanahan:2018pib}.
In principle, a series of moments can be used to reconstruct the PDF.
Although there are calculations of the first moment of the pion gluon PDF, there is little chance that sufficient higher moments of the pion gluon PDF can be obtained to perform such a reconstruction.
A direct lattice calculation of the $x$-dependence of the pion gluon PDF is needed.

In recent years, there has been an increasing number of calculations of $x$-dependent hadron structure in lattice QCD, following the proposal of Large-Momentum Effective Theory (LaMET)~\cite{Ji:2013dva,Ji:2014gla,Ji:2017rah}.
The LaMET method calculates lattice quasi-distribution functions, defined in terms of matrix elements of equal-time and spatially separated operators, and then takes the infinite-momentum limit to extract the lightcone distribution.
The quasi-PDF can be related to the $P_z$-independent lightcone PDF through a factorization theorem that factors from it a perturbative matching coefficient with corrections suppressed by the hadron momentum~\cite{Ji:2014gla}.
The factorization can be calculated exactly in perturbation theory~\cite{Ma:2017pxb,Liu:2019urm}.
Many lattice works have been done on nucleon and meson PDFs, and generalized parton distributions (GPDs) based on the quasi-PDF approach~\cite{Lin:2013yra,Lin:2014zya,Chen:2016utp,Lin:2017ani,Alexandrou:2015rja,Alexandrou:2016jqi,Alexandrou:2017huk,Chen:2017mzz,Alexandrou:2018pbm,Chen:2018xof,Chen:2018fwa,Alexandrou:2018eet,Lin:2018qky,Fan:2018dxu,Liu:2018hxv,Wang:2019tgg,Lin:2019ocg,Chen:2019lcm,Lin:2020reh,Chai:2020nxw,Bhattacharya:2020cen,Lin:2020ssv,Zhang:2020dkn,Li:2020xml,Fan:2020nzz,Gao:2020ito,Lin:2020fsj,Zhang:2020rsx,Alexandrou:2020qtt,Alexandrou:2020zbe,Lin:2020rxa,Gao:2021hxl,Lin:2020rut}.
Alternative approaches to lightcone PDFs in lattice QCD are ``good lattice cross sections''~\cite{Ma:2017pxb,Bali:2017gfr,Bali:2018spj,Sufian:2019bol,Sufian:2020vzb} and the pseudo-PDF approach~\cite{Orginos:2017kos,Karpie:2017bzm,Karpie:2018zaz,Karpie:2019eiq,Joo:2019jct,Joo:2019bzr,Radyushkin:2018cvn,Zhang:2018ggy,Izubuchi:2018srq,Joo:2020spy,Bhat:2020ktg,Fan:2020cpa,Sufian:2020wcv,Karthik:2021qwz}.
However, LQCD calculations of the $x$ dependence of the pion PDFs have only been done for the valence-quark distribution~\cite{Chen:2018fwa,Sufian:2019bol,Izubuchi:2019lyk,Joo:2019bzr,Sufian:2020vzb,Shugert:2020tgq,Gao:2020ito}.

Only recently have lattice calculations of the gluon PDF become possible, when the necessary one-loop matching relations of the gluon PDF were computed for the pseudo-PDF~\cite{Balitsky:2019krf} and quasi-PDF~\cite{Zhang:2018diq,Wang:2019tgg} approaches.
Both approaches make direct calculation of the $x$ dependence of the pion gluon PDF feasible.
In this work, we apply the pseudo-PDF method by using the ratio renormalization scheme to avoid the difficulty of calculating the gluon renormalization factors.
There is a developed procedure for using the pseudo-PDF method to obtain lightcone PDFs from Ioffe-time distributions (ITDs) by matching through two steps, evolution and scheme conversion~\cite{Radyushkin:2018cvn}.
Using the pseudo-PDF method also allows us to use lattice correlators at all boost momenta at small Ioffe-time.
There have been a number of successful pseudo-PDF calculations of nucleon isovector
PDFs~\cite{Orginos:2017kos,Joo:2019jct,Joo:2020spy,Bhat:2020ktg} and pion valence-quark PDFs~\cite{Joo:2019bzr}.
The earliest calculation was done on a quenched lattice~\cite{Orginos:2017kos}, then the pion masses were set closer to the physical pion mass~\cite{Joo:2019jct,Joo:2019bzr,Joo:2020spy}, and the calculation at physical pion mass was done recently~\cite{Bhat:2020ktg}.
The lattice-calculated PDFs in Refs.~\cite{Joo:2019jct,Joo:2019bzr,Joo:2020spy,Bhat:2020ktg} show good agreement with the global-analysis PDFs.

In this work, we present the first calculation of the full $x$-dependent pion gluon distribution using the pseudo-PDF method from two lattice spacings, 0.12 and 0.15~fm, and three pion masses: 690, 310 and 220~MeV.
The rest of the paper is organized as follows.
In Sec.~\ref{sec:cal-details}, we present the procedure to obtain the lightcone gluon PDF from the reduced ITDs,
the numerical setup of lattice simulation, and how we extracted the reduced ITDs from lattice calculated correlators.
In Sec.~\ref{sec:results}, the final determination of the pion gluon PDF from our lattice calculations is compared with the NLO xFitter~\cite{Novikov:2020snp} and JAM pion gluon PDFs~\cite{Barry:2018ort,Cao:2021aci}.
The systematics induced by different steps are studied, and the lattice-spacing and pion-mass dependence are investigated.

\section{Gluon PDF from Lattice Calculation Using Pseudo-PDF Method}\label{sec:cal-details}

In this work, we use the unpolarized gluon operator defined in Ref.~\cite{Balitsky:2019krf},
\begin{equation}\label{eq:gluon_operator}
 {\cal O(z)}\equiv\sum_{i\neq z,t}{\cal O}(F^{ti},F^{ti};z)-\sum_{i,j\neq z,t}{\cal O}(F^{ij},F^{ij};z),
\end{equation}
where the operator ${\cal O}(F^{\mu\nu}, F^{\alpha\beta};z) = F^\mu_\nu(z)U(z,0)F^{\alpha}_{\beta}(0)$, $z$ is the Wilson link length, and the field strength $F_{\mu\nu}$ is defined as
\begin{equation}\label{}
 F_{\mu\nu} = \frac{i}{8a^2g_0}\left(\mathcal{P}_{[\mu,\nu]}+\mathcal{P}_{[\nu,-\mu]}+\mathcal{P}_{[-\mu,-\nu]}+\mathcal{P}_{[-\nu,\mu]}\right),
\end{equation}
where the $a$ is the lattice spacing,
$g_0$ is the strong coupling constant,
and the plaquette $\mathcal{P_{\mu,\nu}}=U_{\mu}(x)U_{\nu}(x+a\hat{\mu})U^{\dag}_{\mu}(x+a\hat{\nu})U^{\dag}_{\nu}(x)$ and $\mathcal{P_{[\mu,\nu]}}=\mathcal{P}_{\mu,\nu}-\mathcal{P}_{\nu,\mu}$.
There is an alternative operator $\sum_{i\neq z,t}{\cal O}(F^{ti},F^{zi};z)$ corresponding to the same matching kernel in Ref.~\cite{Balitsky:2019krf}.
We do not choose this operator, because it vanishes at $P_z=0$ for kinematic reasons, bringing additional difficulties in obtaining the distributions.

Using this operator in Eq.~\ref{eq:gluon_operator}, we calculate lattice gluon matrix elements of the ground-state meson $|0 (P_z) \rangle$ with various boost momenta $P_z$ and Wilson-line displacement lengths $z$.
We then study their dependence on Ioffe time $\nu=zP_z$,
\begin{equation}
\mathcal{M}(\nu,z^2) = \langle 0 (P_z)|{\cal O}(z)|0 (P_z)\rangle,
\label{eq:ME_unpol}
\end{equation}
calling $\mathcal{M}(\nu,z^2)$ the Ioffe-time distribution (ITD).
To eliminate the ultraviolet divergences in the ITD, we construct the reduced ITD (RITD) by taking the ratio of the ITD to the corresponding $z$-dependent matrix element at $P_z=0$, and further normalize the ratio by the matrix element at $z^2=0$ as done in the first quark pseudo-PDF calculation~\cite{Orginos:2017kos},
\begin{equation}
\mathscr{M}(\nu,z^2)=
\frac{\mathcal{M}(zP_z,z^2)/\mathcal{M}(0\cdot P_z,0)}{\mathcal{M}(z\cdot 0,z^2)/\mathcal{M}(0\cdot 0,0)}.
\label{eq:RITD}
\end{equation}
The renormalization of ${\cal O}(z)$ and kinematic factors are cancelled out in the RITDs.
The RITD double ratios used here are automatically normalized to one at $z=0$.

The RITDs are related to the pion gluon $g$ and quark $q_S$ PDFs via the pseudo-PDF matching condition~\cite{Balitsky:2019krf}
\begin{align}
\mathscr{M}(\nu,z^2)&=\int_0^1 dx \frac{xg(x,\mu^2)}{\langle x \rangle_g}R_{gg}(x\nu,z^2\mu^2) \nonumber \\
&+\frac{P_z}{P_0}\int_0^1 dx \frac{xq_S(x,\mu^2)}{\langle x \rangle_g}R_{gq}(x\nu,z^2\mu^2),
\label{matching-eq}
\end{align}
where $\mu$ is the renormalization scale in $\overline{\text{MS}}$ scheme
and $\langle x \rangle_g=\int_0^1 dx \, x g(x,\mu^2)$ is the gluon momentum fraction of the pion.
We can split the gluon-in-gluon $R_{gg}$ contribution in Eq.~\ref{matching-eq} into two parts, which are introduced in Eqs.~21 and 24 in Ref.~\cite{Radyushkin:2018cvn},
\begin{align}
&R_{gg}(y,z^2\mu^2)=R_1(y,z^2\mu^2)+R_2(y), \label{matching-kernel}\\
&R_1(y,z^2\mu^2)=-\frac{\alpha_s(\mu)}{2\pi}N_c\ln\left(z^2\mu^2\frac{e^{2\gamma_E+1}}{4}\right)R_B(y), \label{matching-evolve}\\
&R_2(y)=\cos y-\frac{\alpha_s(\mu)}{2\pi}N_c\left( 2R_B(y)+R_L(y)+R_C(y)\right),
\label{matching-scheme}
\end{align}
where $R_1(y,z^2\mu^2)$ is the term related to evolution,
$R_2(y)$
is the term related to scheme conversion,
$\alpha_s$ is the strong coupling at scale $\mu$,
$N_c=3$ is the number of colors,
and $\gamma_E=0.5772$ is the Euler-Mascheroni constant.
The $z$ in $R_1(y,z^2\mu^2)$ is chosen to be $2e^{-\gamma_E-1/2}/\mu$ so that the log term vanishes, suppressing residuals that contain higher orders of the log term, as discussed in the paper on the one-loop evolution of the pseudo-PDF~\cite{Radyushkin:2018cvn}.
The gluon-in-quark kernel $R_{gq}(y,z^2\mu^2)$, along with $R_B(y)$, $R_L(y)$ and $R_C(y)$, are defined in Eqs.~7.21--23 and the paragraph below Eq.~7.23 in Ref.~\cite{Balitsky:2019krf}.

In this work, we first neglect the pion quark PDF, since the total quark PDF is found to be much smaller than the gluon PDF in global fits~\cite{Barry:2018ort,Novikov:2020snp}.
We will later estimate the systematic uncertainty introduced by this assumption.
The gluon evolved ITD (EITD), $G$ is obtained by using the evolution term $R_1(y,z^2\mu^2)$,
\begin{align}
G(\nu,\mu,z^2)&=\mathscr{M}(\nu,z^2)\nonumber\\
&+\int^1_0 dx\, R_1(x,z^2\mu^2)\mathscr{M}(x\nu,z^2).
\label{evolution}
\end{align}
The $z$ dependence of the EITDs should be compensated by the $\ln{z^2}$ term in the evolution formula.
In principle, the EITD $G$ is free of $z$ dependence and is connected to the lightcone gluon PDF $g(x,\mu^2)$ through the scheme-conversion term $R_2(y)$,
\begin{equation}
G(\nu,\mu)=\int_0^1 dx\, \frac{xg(x,\mu^2)}{\langle x \rangle_g} R_2(x\nu),
\label{conversion}
\end{equation}
so the gluon PDF $g(x,\mu^2)$ can be extracted by inverting this equation.

On the lattice, we use clover valence fermions on three ensembles with $N_f = 2+1+1$ highly improved staggered quarks (HISQ)~\cite{Follana:2006rc} generated by the MILC Collaboration~\cite{Bazavov:2012xda} with two different lattice spacings ($a\approx 0.12$ and 0.15~fm) and three pion masses (220, 310, 690~MeV).
The masses of the clover quarks are tuned to reproduce the lightest light and strange sea pseudoscalar meson masses used by PNDME Collaboration~\cite{Rajan:2017lxk,Bhattacharya:2015wna,Bhattacharya:2015esa,Bhattacharya:2013ehc}.
We use five HYP-smearing~\cite{Hasenfratz:2001hp} steps on the gluon loops to reduce the statistical uncertainties, as studied in Ref.~\cite{Fan:2018dxu}.
We use Gaussian momentum smearing for the quark fields~\cite{Bali:2016lva} to reach higher meson boost momenta.
Table~\ref{table-data} gives the lattice spacing $a$, valence pion mass $M_\pi^\text{val}$ and $\eta_s$ mass $M_{\eta_s}^\text{val}$, lattice size $L^3\times T$, number of configurations $N_\text{cfg}$, number of total two-point correlator measurements $N_\text{meas}^\text{2pt}$, and separation time $t_\text{sep}$ used in the three-point correlator fits for the three ensembles.
This allows us to reach the continuum limit and physical pion mass through extrapolation.
The total amount of measurements vary in $10^5$--$10^6$ for different ensembles.

\begin{table}[!htbp]
\centering
\begin{tabular}{|c|c|c|c|}
\hline
  ensemble  & a12m220 & a12m310 & a15m310 \\
\hline
  $a$ (fm)  & $0.1184(10)$ & $0.1207(11)$  & $0.1510(20)$ \\
\hline
  $M_{\pi}^\text{val}$ (MeV)  & $226.6(3)$ & $311.1(6)$ & $319.1(31)$\\
\hline
  $M_{\eta_s}^\text{val}$ (MeV)  & $696.9(2)$ & $684.1(6)$ & $687.3(13)$\\
\hline
  $L^3\times T$  & $32^3\times 64$ & $24^3\times 64$ & $16^3\times 48$ \\
\hline
  $P_z$ (GeV)  & $[0,2.29]$ &  $[0,2.14]$ &  $[0,2.05]$ \\
\hline
  $N_\text{cfg}$  & 957 & 1013 & 900 \\
\hline
  $N_\text{meas}^\text{2pt}$  &  731,200  & 324,160 & 21,600 \\
\hline
  $t_\text{sep}$  & \{5,6,7,8,9\}   & \{5,6,7,8,9\} & \{4,5,6,7\} \\
\hline
\end{tabular}
\caption{
Lattice spacing $a$, valence pion mass $M_\pi^\text{val}$ and $\eta_s$ mass $M_{\eta_s}^\text{val}$, lattice size $L^3\times T$, number of configurations $N_\text{cfg}$, number of total two-point correlator measurements $N_\text{meas}^\text{2pt}$, and separation times $t_\text{sep}$ used in the three-point correlator fits of $N_f=2+1+1$ clover valence fermions on HISQ ensembles generated by MILC Collaboration and analyzed in this study.
}
\label{table-data}
\end{table}

The two-point correlator for a meson $\Phi$ is
\begin{align}\label{}
C_\Phi^\text{2pt}(P_z;t)&=
 \int dy^3 e^{-i y\cdot P_z} \langle \chi_\Phi(\vec{y},t)|\chi_\Phi(\vec{0},0)\rangle \nonumber \\
  &= |A_{\Phi,0}|^2 e^{-E_{\Phi,0}t} + |A_{\Phi,1}|^2 e^{-E_{\Phi,1}t} + ...,
\label{eq:2pt_fit_formula}
\end{align}
where $P_z$ is the meson momentum in the $z$-direction, $\chi_\Phi=\bar{q}_1\gamma_5 q_2$ is the pseudoscalar-meson interpolation operator,
$t$ is the Euclidean time,
and $|A_{\Phi,i}|^2$ and $E_{\Phi,i}$ are the amplitude and energy for the ground-state ($i=0$) and the first excited state ($i=1$), respectively.

\begin{figure*}[htbp]
\centering
\centering
\includegraphics[width=0.9\textwidth]{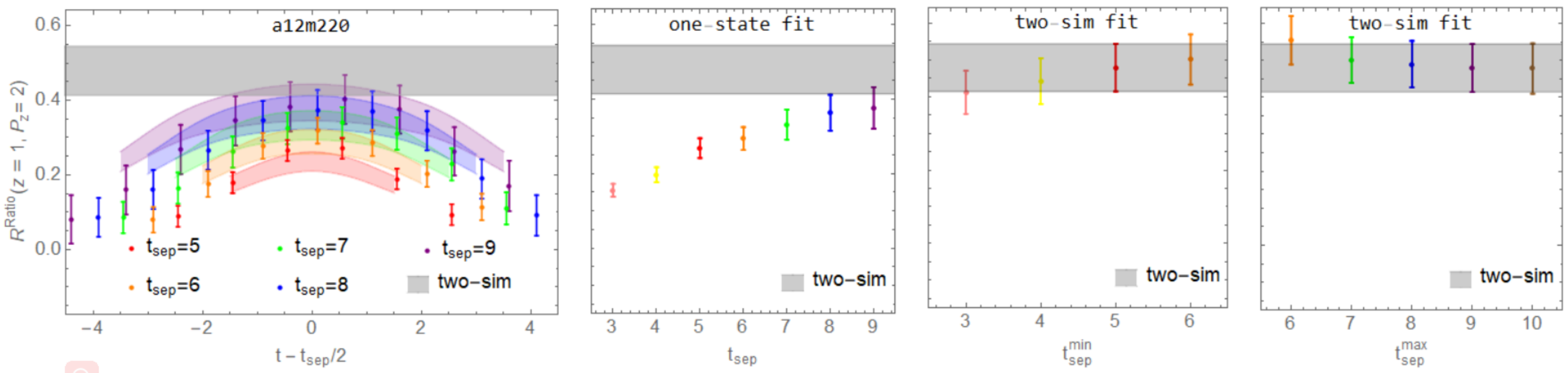}
\includegraphics[width=0.9\textwidth]{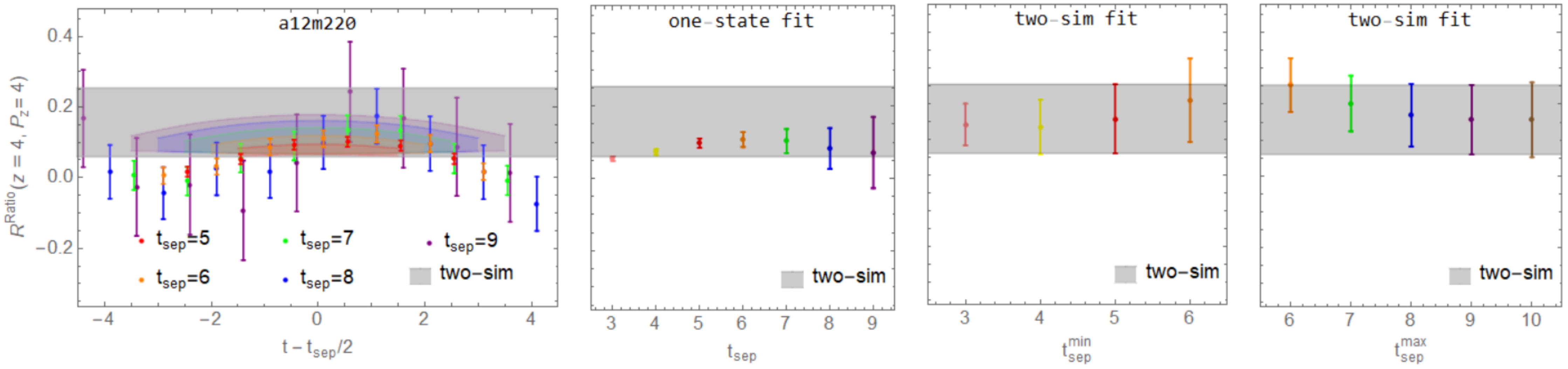}
\caption{
Example ratio plots (left), one-state fits (second column) and two-sim fits (last 2 columns) from the lightest pion mass $a\approx 0.12$~fm, $M_\pi\approx 220$~MeV for $P_z=2\times 2\pi/L$, $z=1$ (upper row) and $P_z=4\times 2\pi/L$, $z=4$ (lower row).
The gray band shown on all plots is the extracted ground-state matrix element from the two-sim fit using $t_\text{sep}\in[5,9]$.
From left to right, the columns are:
the ratio of the three-point to two-point correlators with the reconstructed fit bands from the two-sim fit using $t_\text{sep}\in [5,9]$, shown as functions of $t-t_\text{sep}/2$,
the one-state fit results for the three-point correlators at each $t_\text{sep}\in[3,9]$,
the two-sim fit results using $t_\text{sep}\in[t_\text{sep}^\text{min},9]$ as functions of $t_\text{sep}^\text{min}$, and
the two-sim fit results using $t_\text{sep}\in[5,t_\text{sep}^\text{max}]$
as functions of $t_\text{sep}^\text{max}$.}
\label{fig:Ratio-fitcomp}
\end{figure*}

The three-point gluon correlators are obtained by combining the gluon loop with pion two-point correlators.
The matrix elements of the gluon operators can be obtained by fitting the three-point correlators to the energy-eigenstate expansion,
\begin{align}\label{eq:3ptC}
&C_\Phi^\text{3pt}(z,P_z;t_\text{sep},t) \nonumber \\
&=\int d^3y\, e^{-iy\cdot P_z}\langle \chi_\Phi(\vec y,t_\text{sep})|{\cal O}(z,t)|\chi_\Phi(\vec 0,0)\rangle \nonumber \\
&= |A_{\Phi,0}|^2\langle 0|{\cal O}|0\rangle e^{-E_{\Phi,0}t_\text{sep}} \nonumber \\
&+ |A_{\Phi,0}||A_{\Phi,1}|\langle 0|{\cal O}|1\rangle e^{-E_{\Phi,1}(t_\text{sep}-t)}e^{-E_{\Phi,0}t} \nonumber \\
&+ |A_{\Phi,0}||A_{\Phi,1}|\langle 1|{\cal O}|0\rangle e^{-E_{\Phi,0}(t_\text{sep}-t)}e^{-E_{\Phi,1}t} \nonumber \\
&+ |A_{\Phi,1}|^2\langle 1|{\cal O}|1\rangle e^{-E_{\Phi,1}t_\text{sep}}+ ...,
\end{align}
where $t_\text{sep}$ is the source-sink time separation,
and $t$ is the gluon-operator insertion time.
The amplitudes and energies, $A_{\Phi,0}$, $A_{\Phi,1}$, $E_{\Phi,0}$ and $E_{\Phi,1}$, are obtained from the two-state fits of the two-point correlators.
$\langle 0|{\cal O}|0\rangle$, $\langle 0|{\cal O}|1\rangle$ ($\langle 1|{\cal O}|0\rangle$), and $\langle 1|{\cal O}|1\rangle$ are the ground-state matrix element, the ground--excited-state matrix element, and the excited-state matrix element, respectively.
We extract the ground-state matrix element $\langle 0|{\cal O}|0\rangle$ from the two-state fit of the three-point correlators, or a two-state simultaneous ``two-sim'' fit on multiple separation times with the $\langle 0|{\cal O}|0\rangle$, $\langle 0|{\cal O}|1\rangle$ and $\langle 1|{\cal O}|0\rangle$ terms.

To verify that our fitted matrix elements are reliably extracted, we compare to ratios of the three-point to the two-point correlator
\begin{equation}\label{eq:Ratio}
R^\text{ratio}(z,P_z;t_\text{sep},t) = \frac{C^\text{3pt}(z,P_z;t_\text{sep},t)}{C^\text{2pt}(P_z;t_\text{sep})};
\end{equation}
if there were no excited states, the ratio would be the ground-state matrix element.
The left-hand side of Fig.~\ref{fig:Ratio-fitcomp} shows example ratios for the gluon matrix elements from the lightest pion ensemble, a12m220, at selected momenta $P_z$ and Wilson-line length $z$.
We see the ratios increase with increasing source-sink separation going from 0.60 to 1.08~fm.
At large separation, the ratios begin to converge, indicating the neglect of excited states becomes less problematic.
The gray bands indicate the ground-state matrix elements extracted using the two-sim fit to three-point correlators at five $t_\text{sep}$.
The convergence of the fits that neglect excited states can also be seen in second column of Fig.~\ref{fig:Ratio-fitcomp}, where we compare one-state fits from each source-sink separations:
the one-state fit results increase as $t_\text{sep}$ increases, starting to converge at large $t_\text{sep}$ to the two-sim fit results.

The third and fourth columns of Fig.~\ref{fig:Ratio-fitcomp} show two-sim fits using $t_\text{sep}\in[t_\text{sep}^\text{min},9]$
and
$t_\text{sep}\in[5,t_\text{sep}^\text{max}]$
to study how the two-sim ground-state matrix elements depend on the source-sink separations input into fit.
We observe that the matrix elements are consistent with each other within one standard deviation, showing consistent extraction of the ground-state matrix element, though the statistical errors are larger than those of the one-state fits.
We observe larger fluctuations in the matrix element extractions when small $t_\text{sep}^\text{min}=3$ and 4, or small $t_\text{sep}^\text{max}=6$ and 7, are used. The ground state matrix element extracted from two-sim fits becomes very stable when $t_\text{sep}^\text{min}>4$ and $t_\text{sep}^\text{max}>7$.

\begin{figure}[htbp]
\centering
	\centering
	\includegraphics[width=0.43\textwidth]{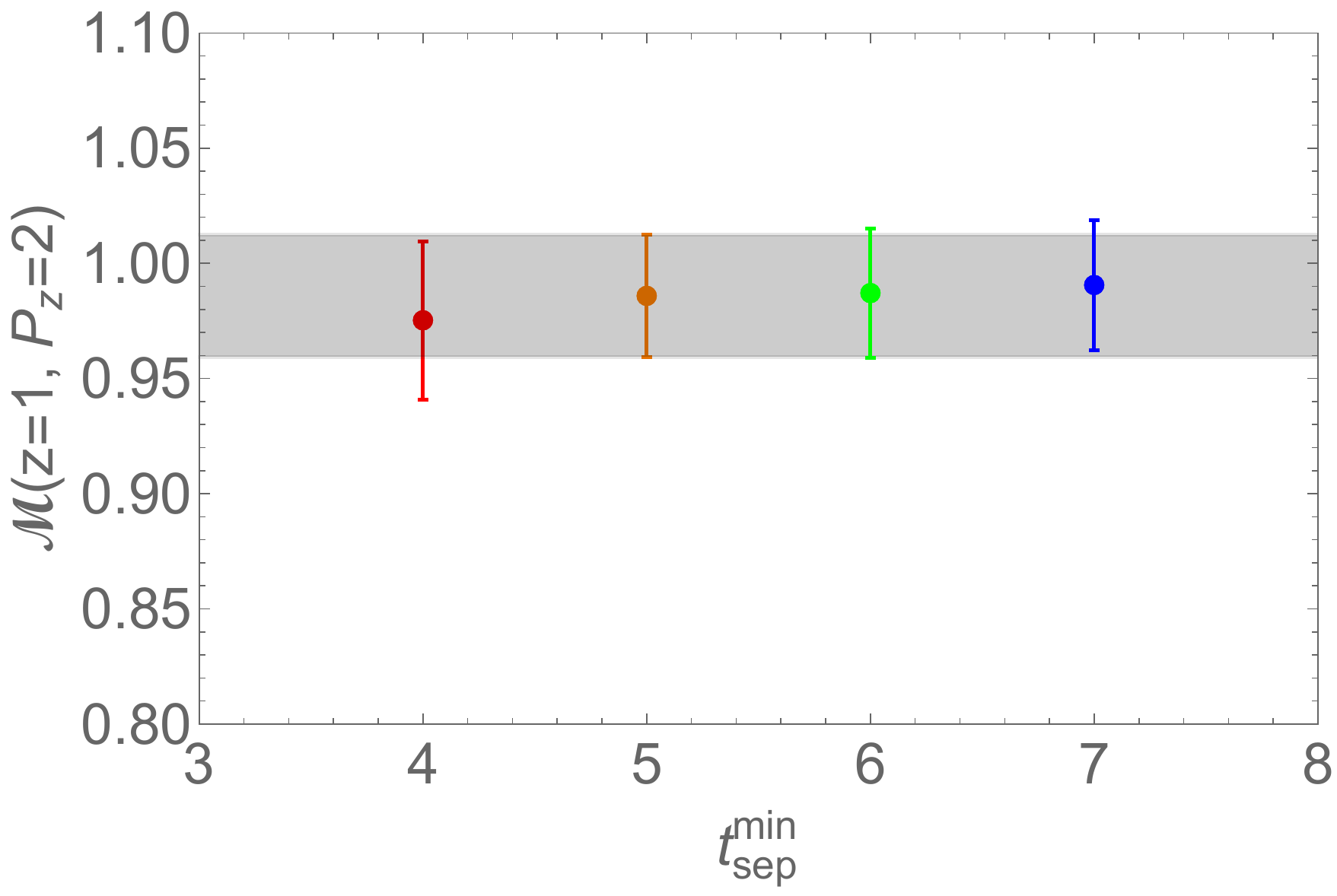}
	\includegraphics[width=0.43\textwidth]{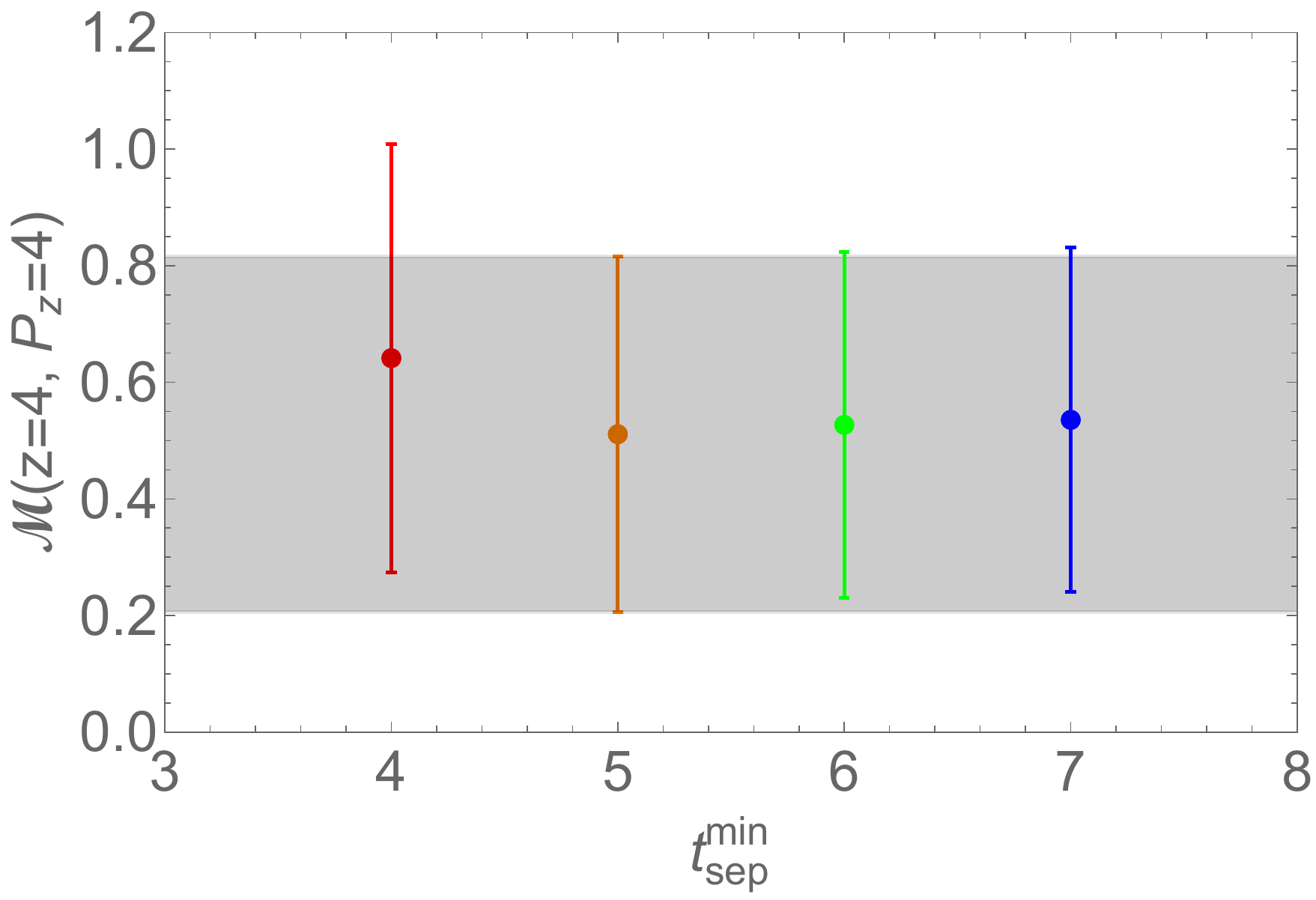}
\caption{
Example RITDs from the a12m220 ensemble as functions of $t_\text{sep}^\text{min}$ for $P_z=2\times 2\pi/L$, $z=1$ (top) and $P_z=4\times 2\pi/L$, $z=4$ (bottom).
The two-sim fit RITD results using $t_\text{sep}\in[t_\text{sep}^\text{min},9]$ are consistent with the ones final chosen $t_\text{sep}\in[5,9]$.}
\label{fig:RITD-fitcomp}
\end{figure}

Figure~\ref{fig:RITD-fitcomp} shows the RITD of the same examples
$P_z=2\times 2\pi/L$, $z=1$ and $P_z=4\times 2\pi/L$, $z=4$ from two-sim fit results using $t_\text{sep}\in[t_\text{sep}^\text{min},9]$.
The RITD results, which are constructed to suppress lattice fluctuations, are very stable over the range of different fits considered.
For a12m310 and a15m310 ensembles, the $t_\text{sep}$ dependence of RITDs is milder than those from a12m220 ensemble due to the heavier pion mass.
Overall, our ground-state RITDs from the two-sim fit are stable, and we use them to extract the gluon PDF.

\section{Results and Discussions}\label{sec:results}

\begin{figure}[htbp]
\centering
	\centering
	\includegraphics[width=0.45\textwidth]{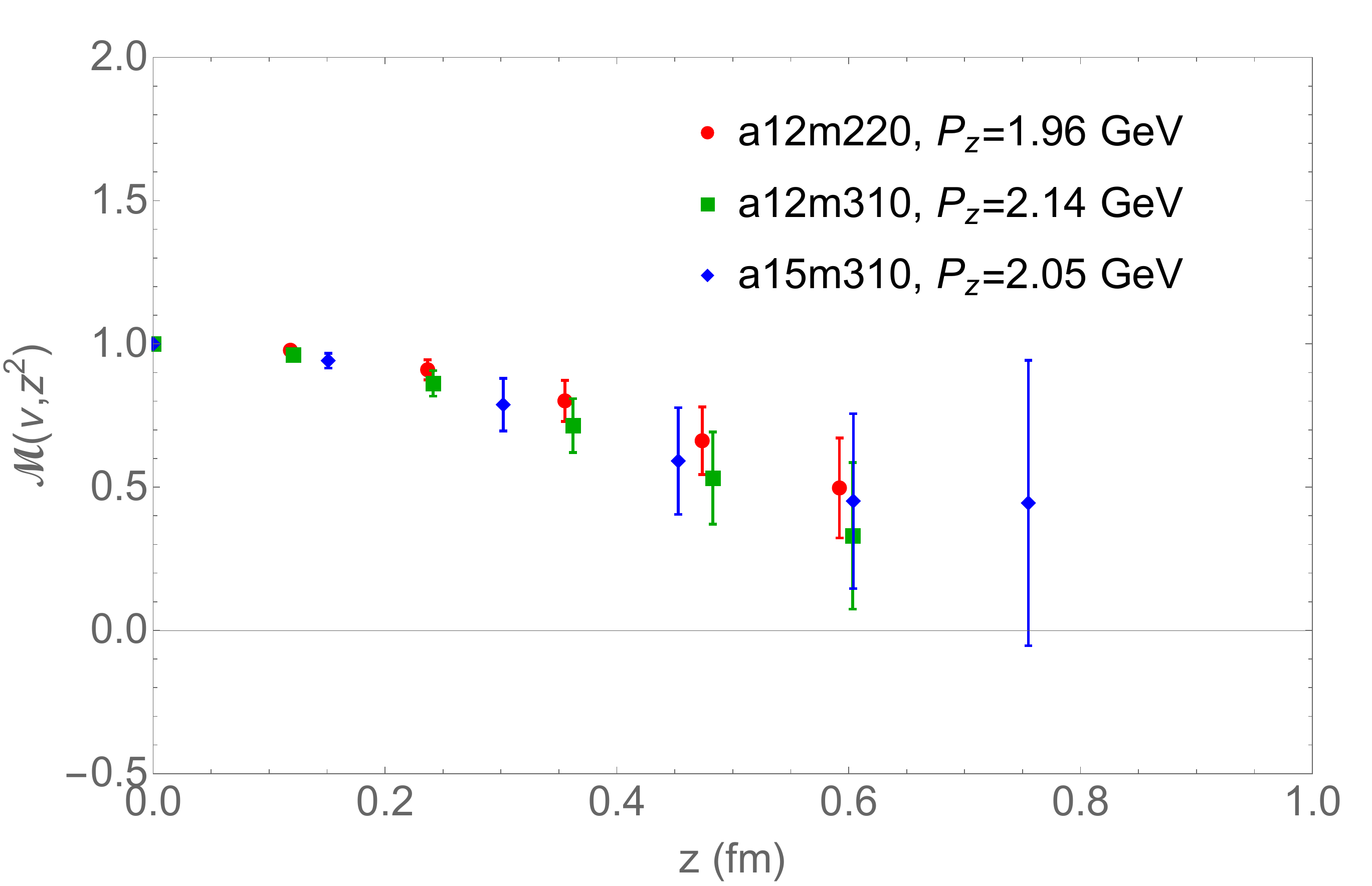}
	\includegraphics[width=0.45\textwidth]{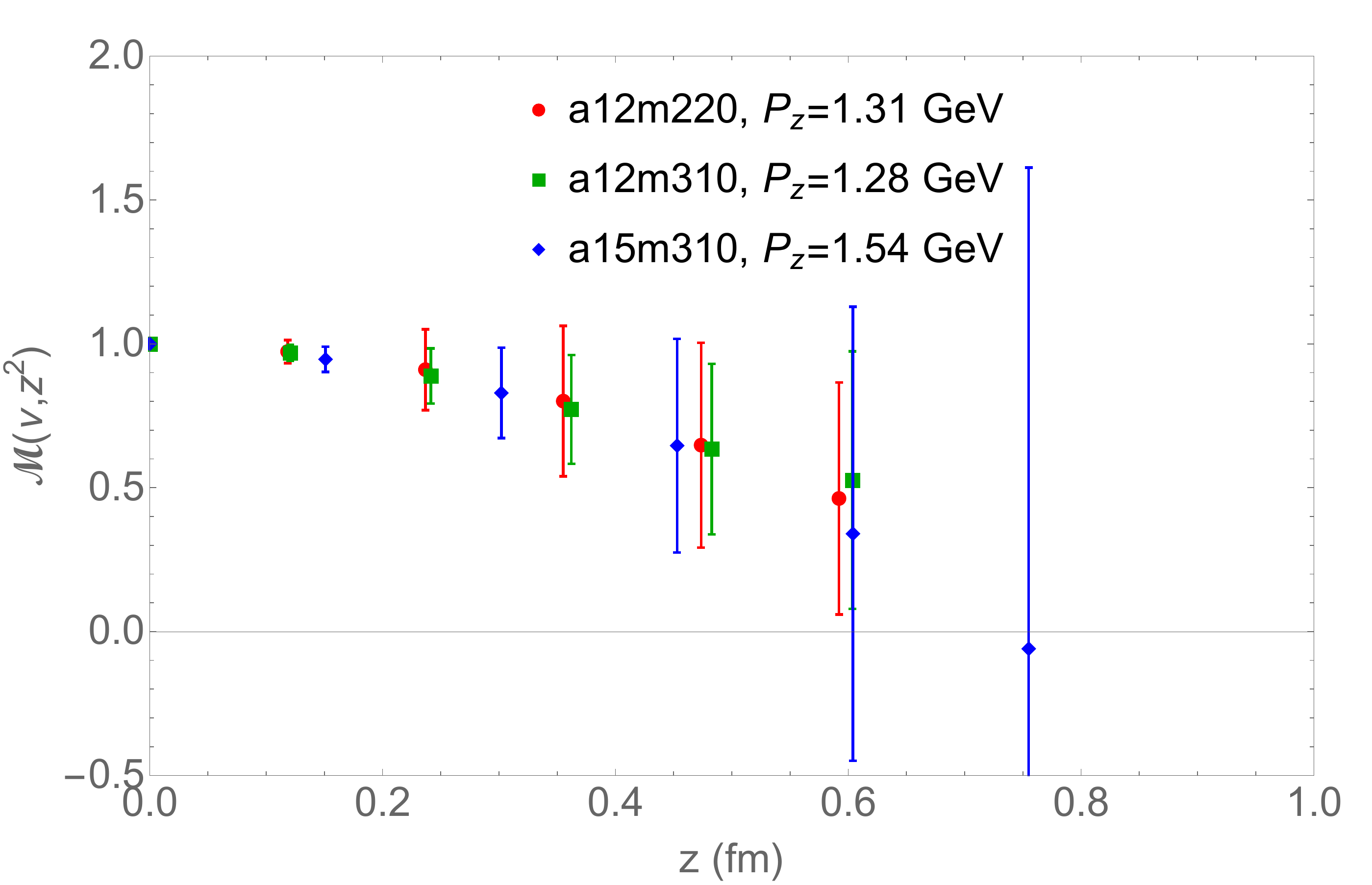}
\caption{The $\eta_s$ (top) and pion (bottom) RITDs at boost momenta $P_z \approx 2$~GeV and 1.3~GeV, respectively, for the a12m220, a12m310, and a15m310 ensembles.
In both cases, we observe weak lattice-spacing and pion-mass dependence.
}
\label{fig:RITD-ensembles}
\end{figure}

Using the RITDs extracted in the previous section, we examine the pion-mass and lattice-spacing dependence.
The top of Fig.~\ref{fig:RITD-ensembles} shows the $\eta_s$ RITDs at boost momentum around 2~GeV as functions of the Wilson-line length $z$ for the a12m220, a12m310, and a15m310 ensembles.
We see no noticeable lattice-spacing dependence.
The bottom of Fig.~\ref{fig:RITD-ensembles} shows the pion RITDs with boost momentum around 1.3~GeV for the same ensembles.
Again, there is no visible lattice-spacing or pion-mass dependence.

\begin{figure}[htbp]
\centering
	\centering
	\includegraphics[width=0.45\textwidth]{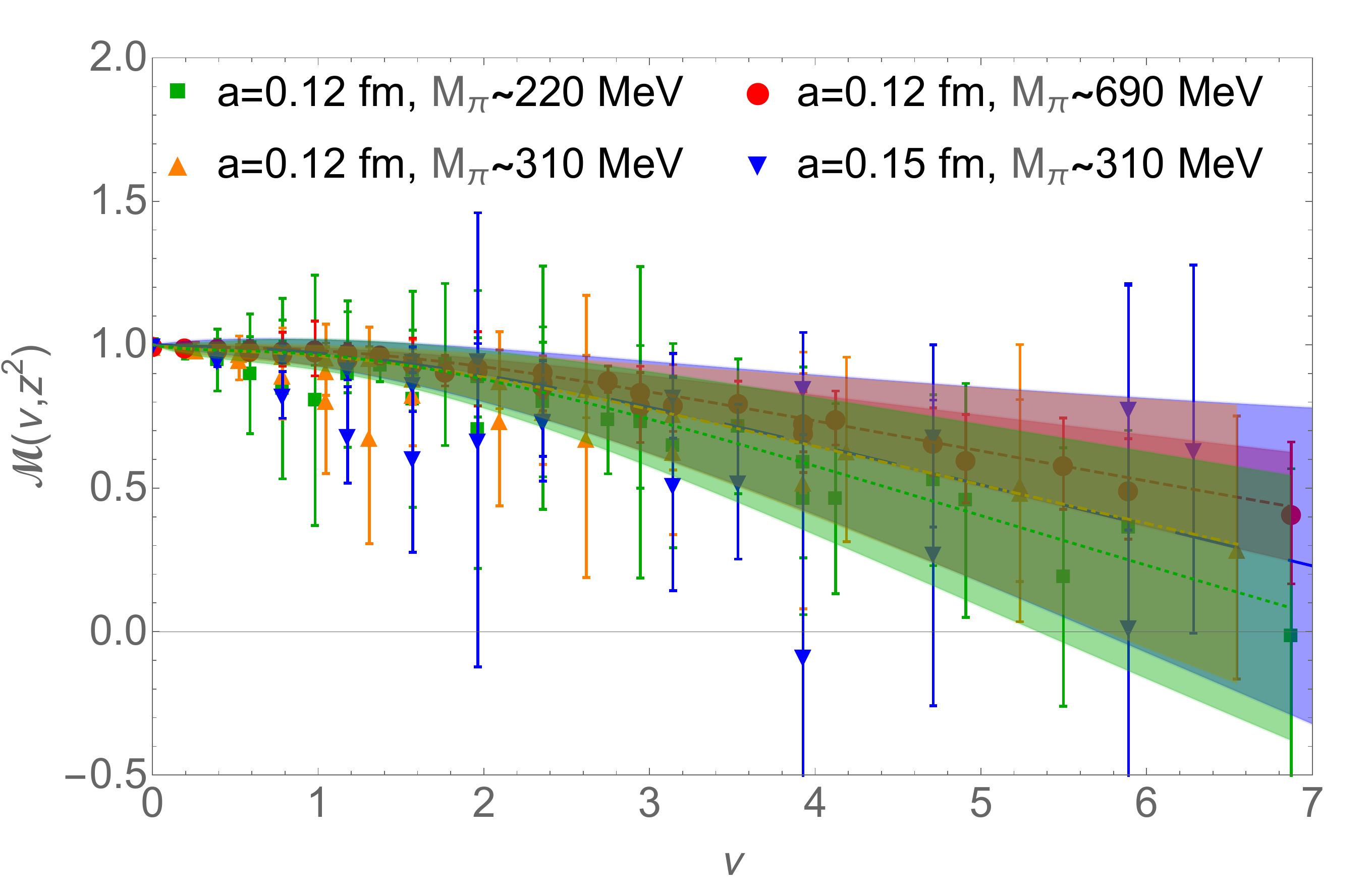}
	\includegraphics[width=0.45\textwidth]{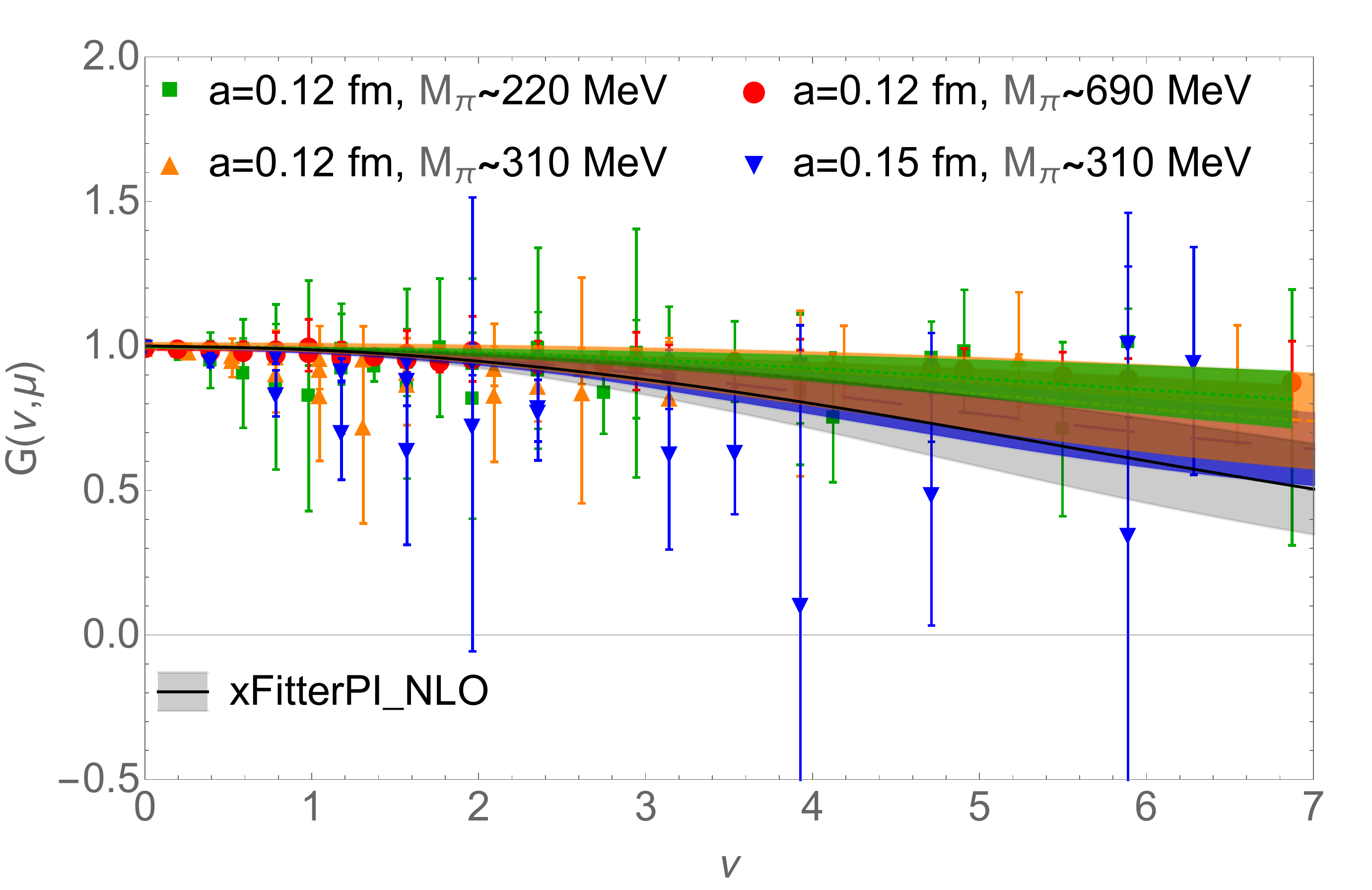}
\caption{
The RITDs $\mathscr{M}$ with reconstructed bands from ``$z$-expansion'' fits (top) and the EITDs $G$ with reconstructed bands from fits (bottom) calculated on ensembles with lattice spacing $a\approx 0.12$~fm, pion masses $M_\pi\approx\{220,310,690\}$~MeV, and $a\approx 0.15$~fm, $M_\pi\approx 310$~MeV, noticing that $a\approx 0.12$, $M_\pi\approx 690$~MeV results are from a12m220 ensemble here.
}
\label{fig:ITD-comparison}
\end{figure}

To extract gluon PDFs, we follow the steps in Sec.~\ref{sec:cal-details} between Eq.~\ref{eq:ME_unpol} and Eq.~\ref{evolution} by first obtaining EITDs and using Eq.~\ref{conversion} to extract $g(x)$.
To obtain EITDs, we need the RITD $\mathscr{M}(\nu,z^2)$ to be a continuous function of $\nu$ to evaluate the $x\in[0,1]$ integral in Eq.~\ref{evolution}.
We achieve this by using a ``$z$-expansion''\footnote{Note that the $z$ in the ``$z$-expansion'' is not related to the Wilson link length $z$ we use elsewhere.} fit~\cite{Boyd:1994tt,Bourrely:2008za} to the RITD.
The following form is used \cite{Joo:2019bzr}:
\begin{equation}
\mathscr{M}(\nu, z^2,M_{\pi}) = \sum_{k=0}^{k_\text{max}}\lambda_k\tau^k,
\label{zexpansion}
\end{equation}
where $\tau=\frac{\sqrt{\nu_\text{cut}+\nu}-\sqrt{\nu_\text{cut}}}{\sqrt{\nu_\text{cut}+\nu}+\sqrt{\nu_\text{cut}}}$.
Then, we use the fitted $\mathscr{M}(\nu, z^2)$ in the integral in Eq.~\ref{evolution}.
The $z$-dependence in the $\mathscr{M}(u\nu, z^2)$ term of the evolution function comes from the one-loop matching term, which is a higher-order correction compared to the tree-level term;
thus, the $z$-dependence can be neglected in $\mathscr{M}(\nu, z^2)$.
We adopt as the best value $\nu_\text{cut}=1$, as used in Ref.~\cite{Joo:2019bzr}, but we also vary $\nu_\text{cut}$ in the range $[0.5,2]$, and the results are consistent.
We fix $\lambda_0 = 1$ to enforce the RITD $\mathscr{M}(\nu, z^2)$ in Eq.~\ref{eq:RITD}.
The expansion order $k_\text{max}=3$ is used, because we can fit all the data points of $P_z\in [1,5]\times 2\pi/L$ ($P_z\in [1,7]\times 2\pi/L$ for a12m220 ensemble)
and $z$ up to 0.6~fm
with $\chi^2/\text{dof}<1$ using a 4-term $z$-expansion for each ensemble.
The reconstructed bands from ``$z$-expansion'' on RITDs are shown in the upper plot in Fig.~\ref{fig:ITD-comparison}.
They describe the RITD data points well for all ensembles.

After we have the continuous-$\nu$ fitted RITDs, we obtain the EITDs through Eq.~\ref{evolution}.
The RITDs $\mathscr{M}$ and EITDs $G$ as functions of $\nu$ on all ensembles studied in this work are shown in Fig.~\ref{fig:ITD-comparison}.
At some $\nu$ values, there are multiple $z$ and $P_z$ combinations for a fixed $\nu$ value.
Therefore, there are points in the same color and symbol overlapping at the same $\nu$ from the same lattice spacing and pion mass.
To match with the lightcone gluon PDF through Eq.~\ref{conversion}, the EITDs $G(\nu,\mu)$ should be free of $z^2$ dependence.
However, the EITDs obtained from Eq.~\ref{evolution} have $z^2$ dependence from neglecting the gluon-in-quark contribution and higher-order terms in the matching.
The EITDs also depend on lattice-spacing $a$ and pion-mass $M_\pi$.
Recall that the RITDs show weak dependence on lattice spacing $a$ and pion mass $M_\pi$.
We see that the effects of $a$ and $M_\pi$ dependence on the EITDs are also not large;
the EITD results from different $a$, $M_\pi$ are mostly consistent with each other, as shown in the second row of Fig.~\ref{fig:ITD-comparison}.
We also observe a weak dependence on $z^2$ for the RITDs and EITDs in Fig.~\ref{fig:ITD-comparison}.

The gluon PDF $g(x,\mu^2)$ can now be extracted from the EITDs using Eq.~\ref{conversion}.
We assume a functional form, also used by JAM~\cite{Barry:2018ort,Cao:2021aci}, for the lightcone PDF to fit the EITD,
\begin{align}
f_g(x,\mu) = \frac{xg(x, \mu)}{\langle x \rangle_g(\mu)} = \frac{x^A(1-x)^C}{B(A+1,C+1)},
\label{functional}
\end{align}
for $x\in[0,1]$ and zero elsewhere.
The beta function $B(A+1,C+1)=\int_0^1 dx\, x^A(1-x)^C$ is used to normalize the area to unity.
Then, we apply the matching formula to obtain the EITD ${G}$ from the functional form PDF using Eq.~\ref{conversion}.
We fit the EITDs $G(\nu,\mu)$ obtained from the parametrization to the EITDs $G(\nu,z^2,\mu,a,M_\pi)$ from the lattice calculation.

\begin{figure}[htbp]
\centering
\centering
\includegraphics[width=0.45\textwidth]{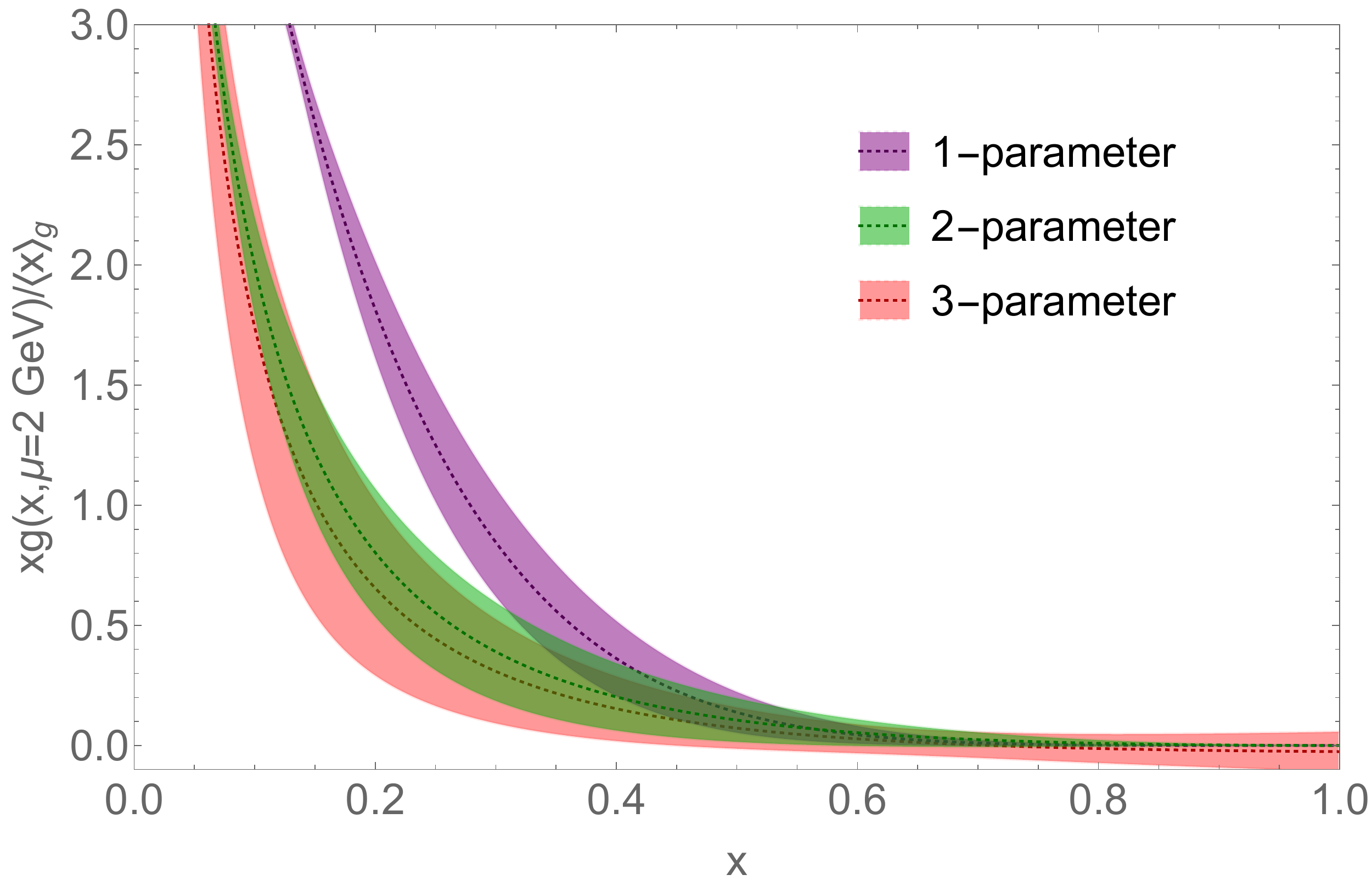}
\caption{
The $xg(x, \mu)/\langle x \rangle_g$ at $\mu^2=4\text{ GeV}^2$ as function of $x$ (bottom) calculated with lattice spacing $a\approx 0.12$~fm, pion masses $M_\pi\approx 220$~MeV with the fitted bands of  $z_\text{max}\approx 0.6$~fm from the 1-, 2- and 3-parameter fits described in Eq.~\ref{functional} and the paragraph after it.
}
\label{fig:fitform-comparison}
\end{figure}

We investigate the systematic uncertainty introduced by the different parametrization forms which are commonly used for $f_g(x,\mu)$ in PDF global analysis and some lattice calculations.
The first one is the 2-parameter form in Eq.~\ref{functional}.
Second, we consider the 1-parameter form $N_1 (1-x)^C$ used in xFitter's analysis~\cite{Novikov:2020snp} (also used in Ref.~\cite{Aurenche:1989sx,Sutton:1991ay}), which is equivalent to Eq.~\ref{functional} with $A=0$.
Third, we consider a 3-parameter form $N_3 x^A(1-x)^C(1+D\sqrt{x})$.
We fit the three different forms to the EITDs of lattice data with $z_\text{max}\approx 0.6$~fm by applying the scheme conversion Eq.~\ref{conversion} to the 1-, 2- and 3-parameter PDF forms.
Here, we focus on the result from the lightest pion mass $M_{\pi}\approx 220$~MeV at lattice spacing $a\approx 0.12$~fm.
The $\chi^2/\text{dof}$ of the fits decreases as $1.47(72)$, $1.08(68)$, to $1.04(41)$, shows slightly better fit quality for 2- and 3-parameter fits.
As shown in Fig.~\ref{fig:fitform-comparison}, there is a big discrepancy between the $f_g(x,\mu)$ fit bands from the 1-parameter fit and the 2-parameter fit in the $x<0.4$ region, but the discrepancy between the 2- and 3-parameter fits is much smaller.
Therefore, we conclude that 1-parameter fit on lattice data here is not quite reliable, and the fit results converge at the 2- and 3-parameter fits.
The same conclusions hold for all other ensembles and pion masses.
Therefore, using the 2-parameter form defined Eq.~\ref{functional} (same parametrization as JAM) for our final results is very reasonable.

\begin{figure*}[htbp]
\centering
	\centering
	\includegraphics[width=0.45\textwidth]{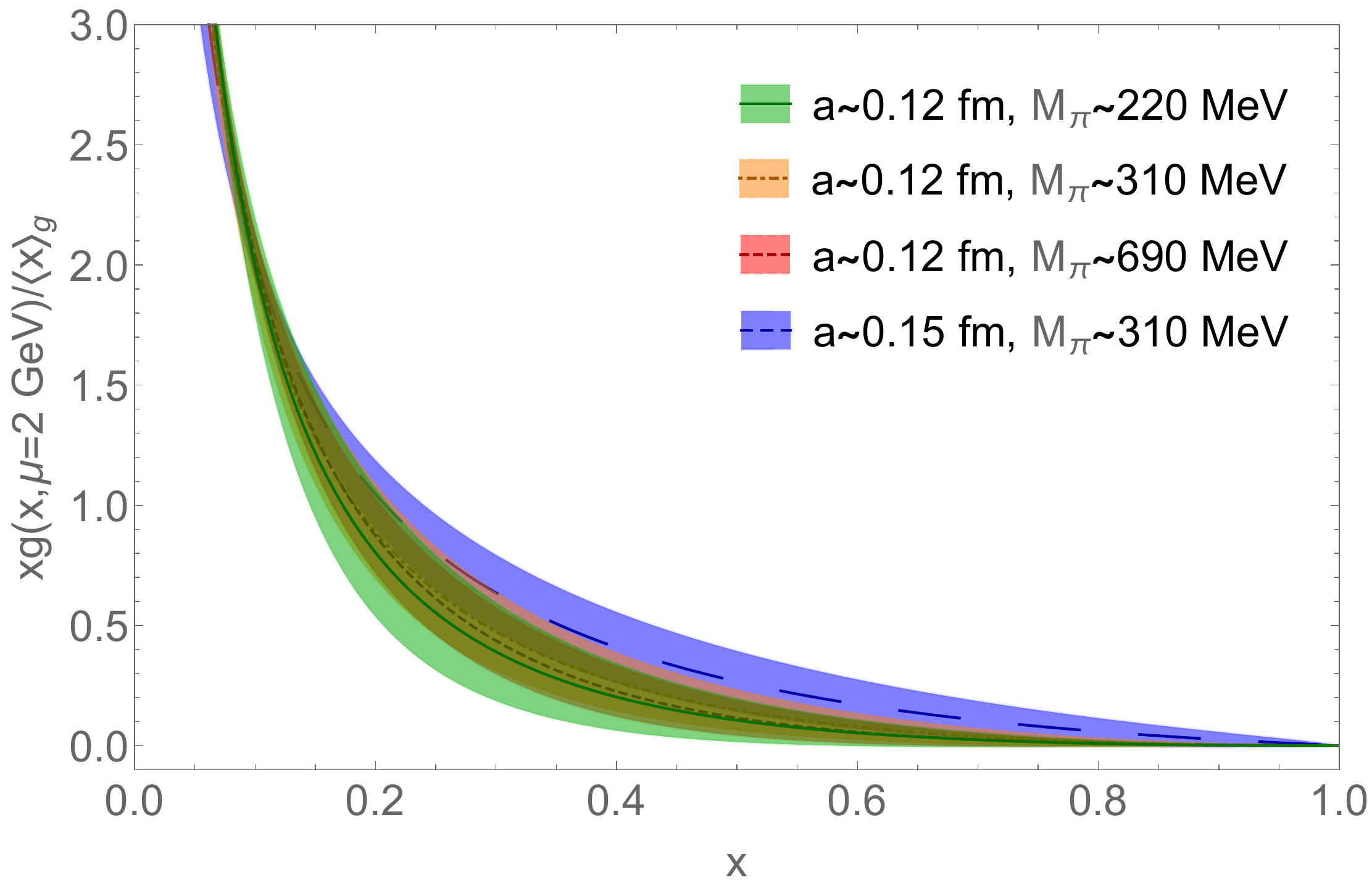}
	\includegraphics[width=0.45\textwidth]{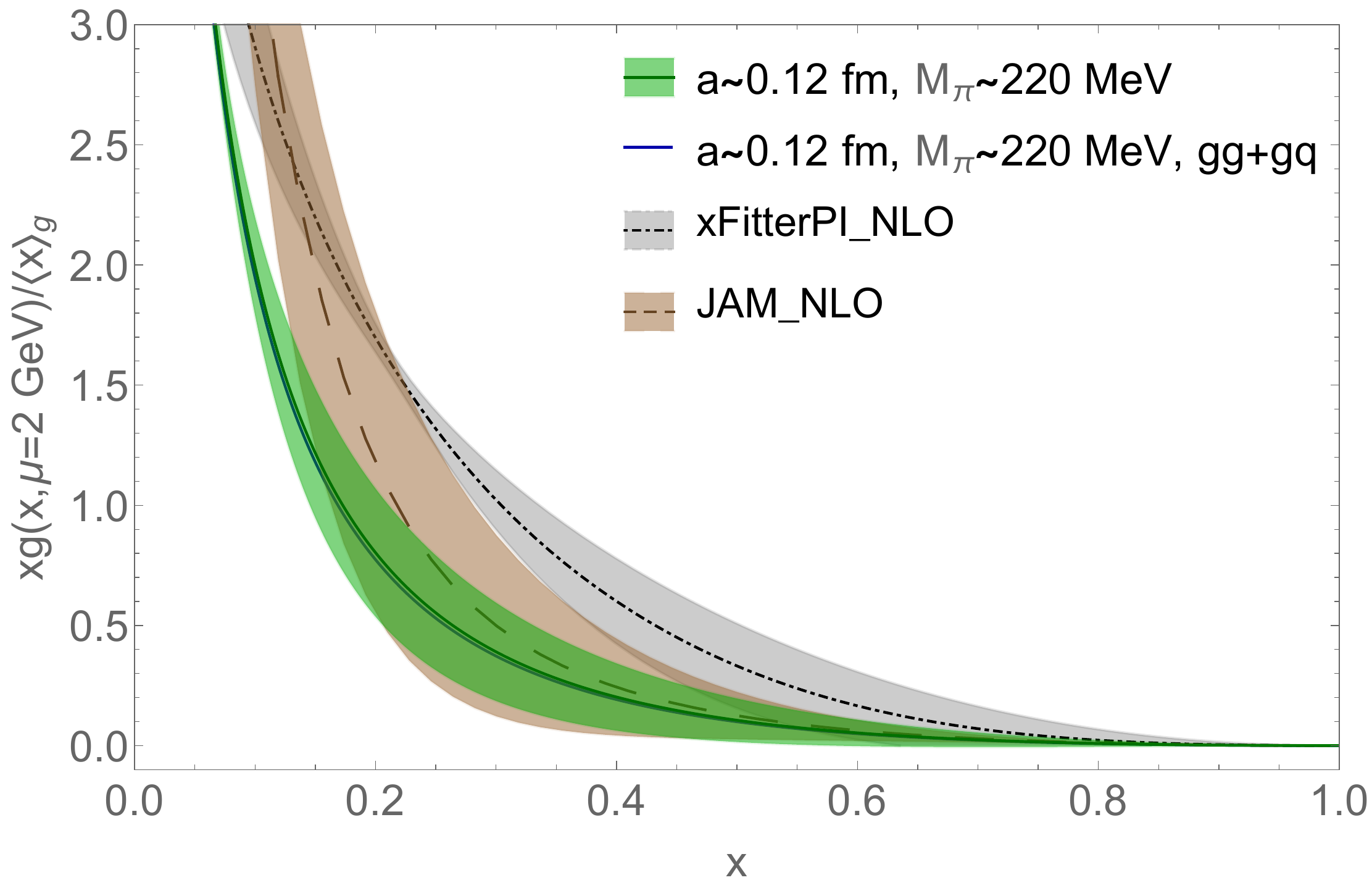}
\caption{
The pion gluon PDF $xg(x, \mu)/\langle x \rangle_g$ as a function of $x$ obtained from the fit to the lattice data on ensembles with lattice spacing $a\approx\{0.12,0.15\}$~fm, pion masses $M_\pi\approx\{220,310,690\}$~MeV (left), and $xg(x, \mu)/\langle x \rangle_g$ as function of $x$ obtained from lattices of $a\approx 0.12$~fm, $M_\pi\approx 220$~MeV (right), compared with the NLO pion gluon PDFs from xFitter and JAM at $\mu=2$~GeV in the $\overline{\text{MS}}$ scheme.
The JAM error shown is overestimated due to lack of available correlated uncertainties in its constituent components.
Our PDF results are consistent with JAM~\cite{Barry:2018ort,Cao:2021aci} for $x>0.2$ and xFitter~\cite{Novikov:2020snp} for $x>0.5$.
}
\label{fig:xgx}
\end{figure*}

Another source of systematic uncertainty comes from neglecting the contribution of the quark term in Eq.~\ref{matching-eq} based on the assumption (motivated by global fits) that the pion $q_S(x)$ is smaller than the gluon PDF.
Currently, there are no $q_S(x)$ results from lattice simulation since only the valence distribution of the pion has been done.
Thus, we estimate the systematic due to omitting the $q_S(x)$ contribution by using the pion quark PDFs from xFitter~\cite{Novikov:2020snp} at NLO.
Using these, we obtain revised RITDs and EITDs including the gluon-in-quark $R_{gq}$ term focusing on example from the $a\approx 0.12$~fm, pion mass $M_\pi\approx 220$~MeV lattice, repeating the same procedure from Eq.~\ref{zexpansion} and fitting the EITDs with Eq.~\ref{functional}.
On the right-hand side of Fig.~\ref{fig:xgx}, we show the mean value of $xg(x, \mu)/\langle x \rangle_g$ with both gluon-in-gluon (gg) and gluon-in-quark (gq) contributions (the blue solid line) compared to the a12m220 results using the gluon-in-gluon contribution only (the green solid line).
There are 5 to $10\%$ differences in the mean value including the gluon-in-gluon contribution for $x<0.9$, which indicates that the gluon-in-quark contribution is relatively small at $\mu^2=4\text{ GeV}^2$ compared to the current statistical errors in the small-$x$ region.
In the $x>0.9$ region, the gluon-in-quark contribution becomes more significant, but it remains smaller than the statistical error.
Once studies are available with sufficiently reduced statistical uncertainty in the large-$x$ region, the quark contribution will need to be included.

From the above analyses of the choice of fit form and the contribution of the quark term, we conclude that these systematics are negligible relative to the current statistics.
Therefore, we adopt the $z_\text{max}\approx 0.6$~fm ($z_\text{max}\approx 0.75$~fm for a15m310 ensembles) fits to the EITDs, neglect the quark contribution term in the matching, and use the Eq.~\ref{conversion} fit form for our final results on all lattice ensembles.
The $xg(x, \mu)/\langle x \rangle_g$ reconstructed fit bands of these ensembles are shown in the left plot in Fig.~\ref{fig:xgx}, comparing results from different lattice spacings and pion masses.
The reconstructed fit bands with different pion mass $M_\pi\approx\{220,310,690\}$~MeV are consistent at the same lattice spacing $a\approx 0.12$~fm, indicating mild gluon PDF dependence on pion mass.
Similarly, when comparing lattice-spacing dependence of pion PDFs using data around pion mass $M_\pi\approx 310$~MeV, we find that fitted PDF is slightly smaller in the $x>0.1$ region for the 0.12-fm lattice, but still within one sigma, which indicates the lattice-spacing dependence is also mild.
We also note that the bands from different ensembles show a differing speed of fall-off as $x\to 1$ in the large-$x$ region.
We study this fall-off behavior in more depth below.

The behavior of the gluon PDF fall-off in the large-$x$ region is widely studied in both theory and global analyses.
Perturbative QCD studies~\cite{Close:1977qx,Sivers:1982wk} and DSE calculations~\cite{Bednar:2018mtf,Freese:2021zne} suggest that the gluon distribution $g(x,\mu^2) \sim (1-x)^C$ with $C\approx 3$ in the limit $x\to 1$.
The prediction from perturbative QCD~\cite{Sivers:1982wk} is based on the idea that the gluon PDF should be suppressed at large $x$ relative to the quark PDF, because the quarks are the sources of large-$x$ gluons;
that is, $g(x,\mu^2)/q_v(x,\mu^2)\to 0$ as $x\to 1$.
Early fits of experimental data gave $C\approx 2$~\cite{Aurenche:1989sx,Sutton:1991ay} or $C<2$~\cite{Gluck:1991ey,Gluck:1999xe}, but the more recent global analysis from JAM collaboration yielded $C>3$~\cite{Barry:2018ort,Cao:2021aci} and xFitter collaboration found $C\approx 3$~\cite{Novikov:2020snp}.
Our fitted parameter $C$ is $3.6(1.5)$, $3.3(2.0)$, $4.7(2.8)$ for $M_\pi\approx\{690,310,220\}$~MeV, respectively, at lattice spacing $a\approx 0.12$~fm.
These $C$ results are consistent with each other and show a slightly increasing trend as the pion mass approaches the physical pion mass.
For lattice spacings $a\approx\{0.15,0.12\}$~fm, $C=\{2.2(1.5),3.3(2.0)\}$, respectively, at $M_\pi\approx 310$~MeV, which suggests that $C$ will increase toward the continuum limit.
We also investigate the effect of the gluon-in-quark contribution on the $C$ value, and it makes about $0.1$ difference, which we neglect.
Given that both the pion-mass and lattice-spacing extrapolations seem to show increasing $C$, it seems reasonable to conclude from this lattice-QCD study that $C>3$.

We compare our reconstructed gluon PDF to those from global fits on the right-hand side of Fig.~\ref{fig:xgx}.
It shows the $xg(x, \mu)/\langle x \rangle_g$ reconstructed fit band of $a\approx 0.12$~fm, $M_\pi\approx 220$~MeV lattice and the NLO pion gluon PDFs from xFitter~\cite{Novikov:2020snp} and JAM~\cite{Barry:2018ort,Cao:2021aci} at $\mu^2=4\text{ GeV}^2$.
The JAM band appears somewhat wider than expected, because we reconstruct it by dividing $xg(x, \mu)$ by the mean value of $\langle x \rangle_g$;
the correlated values needed for a correct error estimation were not available.
Note that xFitter uses the fit form of Eq.~\ref{functional} with $A=0$.
Our fitted pion gluon PDF is consistent with JAM for $x>0.2$ and with xFitter for $x>0.5$ within one sigma.
We see in this comparison that our results are of similar error size as the  global-fit analysis and are useful to provide constraints from theoretical calculation in addition to the experimental data.

\section{Summary}\label{sec:summary}

In this work, we presented the first calculation of the pion gluon PDF from lattice QCD and studied its pion-mass and lattice-spacing dependence using the pseudo-PDF approach.
We employed clover valence fermions on ensembles with $N_f = 2+1+1$ highly improved staggered quarks (HISQ) at two lattice spacings ($a\approx 0.12$ and 0.15~fm) and three pion masses (220, 310 and 690~MeV).
These ensembles allowed us to probe the dependence of the pion gluon PDF on pion mass and lattice spacing.
In both cases, the dependence appears to be weak compared to the current statistical uncertainty.

We investigated the systematics associated with the functional form used in the reconstruction fits as well as the systematics caused by neglecting the quark contribution in the matching.
The effect of the assumed gluon PDF fit form was investigated by using various forms, which are all commonly used or proposed in other PDF works.
We observe large effects changing the fit to $xg(x, \mu)/\langle x \rangle_g$ from 1- to 2-parameter form but convergence at 3 parameters.
This implies the 2-parameter fits are sufficient for our calculation, and our finial pion gluon PDF results are presented using the 2-parameter fit results.
We used the pion quark PDF from xFitter to make an estimation of the quark contribution to the pion gluon RITD.
We found the systematic errors it contributed are smaller than $10\%$ of the statistical errors.

Our pion gluon PDF for the lightest pion mass is consistent with JAM for $x>0.2$ and with xFitter for $x>0.5$ within uncertainty, as shown in our final comparison plots of the pion gluon PDF.
We also studied the asymptotic behavior of the pion gluon PDF in the large-$x$ region in terms of $(1-x)^C$.
$C>3$ is implied from our study at two lattice spacings and three pion masses.
The future study of the pion gluon PDF from the lattice QCD with improved precision and systematic control when combined in global-fit analyses with the results of
anticipated experiments~\cite{Denisov:2018unj,Arrington:2021biu,Anderle:2021wcy,Denisov:2018unj} will provide best determination of the gluon content within the pion.

\section*{Acknowledgments}

We thank MILC Collaboration for sharing the lattices used to perform this study. The LQCD calculations were performed using the Chroma software suite~\cite{Edwards:2004sx}.
We thank JAM Collaboration for providing us the pion $xg(x)$ data with uncertainties for comparison.
ZF thanks Rui Zhang for useful discussions in the earlier stage of this project.
This research used resources of the National Energy Research Scientific Computing Center, a DOE Office of Science User Facility supported by the Office of Science of the U.S. Department of Energy under Contract No. DE-AC02-05CH11231 through ERCAP;
facilities of the USQCD Collaboration, which are funded by the Office of Science of the U.S. Department of Energy,
and supported in part by Michigan State University through computational resources provided by the Institute for Cyber-Enabled Research (iCER).
The work of ZF and HL are partially supported by the US National Science Foundation under grant PHY 1653405 ``CAREER: Constraining Parton Distribution Functions for New-Physics Searches''.

%

\end{document}